\documentclass[aip,reprint,pop,groupedaddress,letterpaper]{revtex4-1}
\usepackage{latexsym, amssymb, amsmath, graphicx, tabularx}

\begin{document}

\title{Experimental characterization of a transition from collisionless to collisional interaction between head-on-merging
supersonic plasma jets} 

\author{A. L. Moser}\thanks{Now at General Atomics, San Diego, CA; electronic mail:  mosera@fusion.gat.com.}
\author{S. C. Hsu}\thanks{Electronic mail: scotthsu@lanl.gov.}
\affiliation{Los Alamos National Laboratory, Los Alamos, New Mexico, 87545, USA}

\date{\today}

\begin{abstract} 
We present results from
experiments on the head-on merging of two supersonic plasma jets in an initially collisionless regime
for the counter-streaming ions.
The plasma jets are of either an argon/impurity or hydrogen/impurity mixture and are produced by pulsed-power-driven railguns.
Based on time- and space-resolved fast-imaging, multi-chord interferometry, and survey-spectroscopy
measurements of the overlapping region between the merging jets,
we observe that the jets initially interpenetrate, consistent with calculated inter-jet ion collision lengths, which are long.
As the jets interpenetrate, a rising 
mean-charge state causes a rapid decrease in the inter-jet ion collision length.   Finally, the interaction becomes collisional and the jets stagnate, eventually producing structures consistent with collisional shocks. These 
experimental observations can
aid in the validation of plasma collisionality and ionization models for plasmas with complex equations of state.
\end{abstract}


\maketitle 

\section{Introduction}

There has been substantial recent interest in the study of colliding plasmas, motivated by a range of research
topics including the potential effects of colliding hohlraum plasmas\cite{bosch92pfb,rancu95prl,wan97pre} on cross-beam
energy transfer\cite{peterson14pop} in inertial confinement fusion,\cite{atzeni04} 
collisional and collisionless-shock studies,\cite{sagdeev91sa,woolsey01pop,romagnani08prl,constantin09ass,kuramitsu11prl,ross12pop,schaeffer12pop,swadling13pop_a,ross13prl,merritt13prl,li13prl,fox13prl,merritt14pop,swadling14prl} and applications such as pulsed laser deposition or laser-induced
breakdown spectroscopy.\cite{luna07jap,sanchez-ake10,al-shboul14pop}
Some of these colliding-plasma interactions can be in a regime that 
is neither purely collisional, in which the plasma can be approximated as a fluid, nor purely collisionless, in which 
classical collisions between particles can be neglected, a situation 
which complicates modeling.\cite{berger91pof,larroche93pfb,rambo94pop,rambo95pop,jones96pop,thoma13pop} 

In this work, we present time- and space-resolved diagnostic measurements
of the overlapping region between two head-on-merging supersonic plasma jets.
From the measurements, we deduce that (a)~the jets initially interpenetrate in
a collisionless manner and then subsequently stagnate and (b) the transition from interpenetration
to stagnation is due to a rising mean-ionization state $\bar{Z}=n_e/n_{tot}$, where $n_{tot}=n_i+n_n$ is the total 
ion-plus-neutral density, that drastically reduces
the inter-jet ion--ion collisional mean free path ($\sim \bar{Z}^{-4}$).

Collisionality estimates based on measured single-jet parameters predict that argon jets would undergo a collisionless interaction (inter-jet ion collision lengths are hundreds of centimeters, much larger than the scale of the 
experiment) and that hydrogen jets would interact collisionally (inter-jet ion collision lengths are a few centimeters).
In the latter case, each jet would act as a barrier for the other, launching collisional shocks.  We show instead that both hydrogen and argon plasmas demonstrate an ionization-mediated transition from collisionless interpenetration
to collisional stagnation, with evidence for collisional shock formation upon stagnation.
The data presented here can aid the validation of fundamental physics models of 
plasma collisionality (e.g., \textcite{jones96jcp}) and ionization (e.g., \textcite{chung05}), especially in plasmas with 
complex equations of state\@.

\section{Overview of the experiment}

\subsection{Experimental apparatus}\label{sec:setup}

\begin{figure}
	\includegraphics[width=\linewidth]{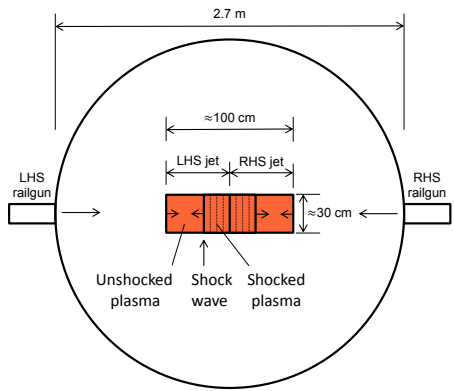}
	\caption{Sketch of head-on merging geometry showing formation of shocks.  \label{fig:geometry}}
	\end{figure}

	These experiments were performed on the Plasma Liner Experiment\cite{hsu12ieee,hsu12pop, hsu14jpp} at Los Alamos National Laboratory. The experiment uses a 2.7-m-diameter spherical vacuum chamber (Fig.~\ref{fig:geometry}), maintained at a background pressure of 
	$\approx5$~$\mu$Torr (neutral density $\sim10^{11}$~cm$^{-3}$).  Plasma jets
	are generated by two pulsed-power-driven railguns,\cite{hsu12pop,hsu14jpp} which can be mounted on any of 60 
	vacuum ports.  For the experiments discussed here, railguns are mounted on two 	
	directly opposed ports to produce the highest relative velocity; gun nozzles face each other and are separated by $\approx220$~cm.  This produces head-on merging 
	between high-Mach-number plasma jets (Fig.~\ref{fig:geometry}), at significantly higher relative velocity (due to geometry) and lower density (due to jet expansion) than our recent work demonstrating formation of collisional shocks in oblique-merging jet experiments,\cite{merritt13prl,merritt14pop} putting these head-on experiments in a less-collisional regime.	
	
	\begin{figure*}
	\includegraphics[width=\linewidth]{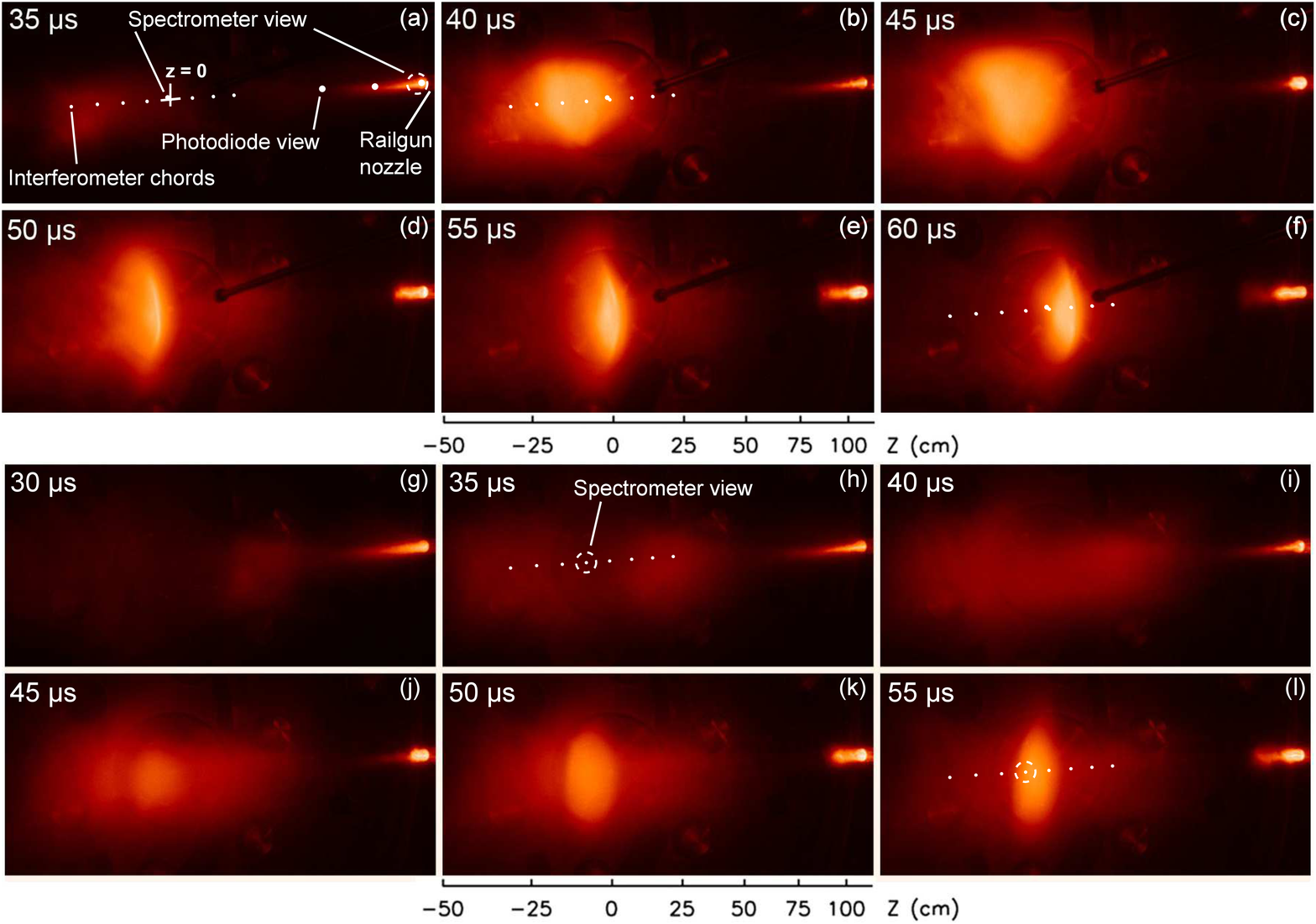}\\
	\caption{Fast-camera images give an overview of the jet-merging interaction.  (a)-(f) Argon jet-merging interaction from $t=35$--60~$\mu$s
	(shots \#1834, 1833, 1836, 1837, 1838, 1845).   (g)-(l) Hydrogen jet-merging interaction from $t=30$--50~$\mu$s
	(shots \#1607, 1610, 1611, 1612, 1614, 1615).  Diagnostic lines-of-sight positions for the photodiode array, spectrometer, and interferometer chords are shown in (a) and chords are noted on images at analysis times.  Images are logarithmically scaled and false-colored.\label{fig:time_series}}
	\end{figure*}

	A DiCam visible-light intensified charge-coupled device (CCD)
	camera (used in single-frame mode, with a 20 ns exposure time), with an $\approx150$ cm field-of-view encompassing chamber center and one plasma railgun nozzle, 
	provides insight into the 
	overall structure and evolution of the plasma jets and the two-jet 
	interaction (Fig.~\ref{fig:time_series}).  
	A three-photodiode array, sensitive to $300$--850~nm light,
	measures plasma emission versus time at 3, 28, and 53~cm from 
	the nozzle  [Fig.~\ref{fig:time_series}(a)], providing information about initial bulk jet velocity via the time-of-flight of the signal peak.  A SpectraPro visible--near-infrared survey spectrometer (spectral resolution $\approx0.152$~nm/pixel, view chord diameter $\approx1.5$ or 7~cm) can be moved between several locations: the gun nozzle [$\approx7$~cm, Fig.~\ref{fig:time_series}(a)], aligned 
with an interferometer chord 7.5 cm from chamber center [$\approx7$~cm, Fig.~\ref{fig:time_series}(h) and (l)], or aligned with an interferometer chord at chamber center [$\approx1.5$~cm, Fig.~\ref{fig:time_series}(a),(b), and (f)].
The spectrometer is used in the iterative data-analysis process described below and provides electron density via Stark broadening of the H-$\beta$ line,\cite{hsu12pop} when present.   
An eight-chord 561-nm laser interferometer\cite{merritt12rsia,merritt12rsib} measures phase shift
$\Delta\Phi$, from the line-integrated effect of free electrons and electrons bound in ions and neutrals, as a function of time.  The chords span the jet-merging region, 
from $z=-30$~cm to $z=22.5$~cm, with a chord separation of 7.5~cm and chord diameter $\approx3$~mm [Fig.~\ref{fig:time_series}(a)].

\begin{figure}
	\includegraphics[width=\linewidth]{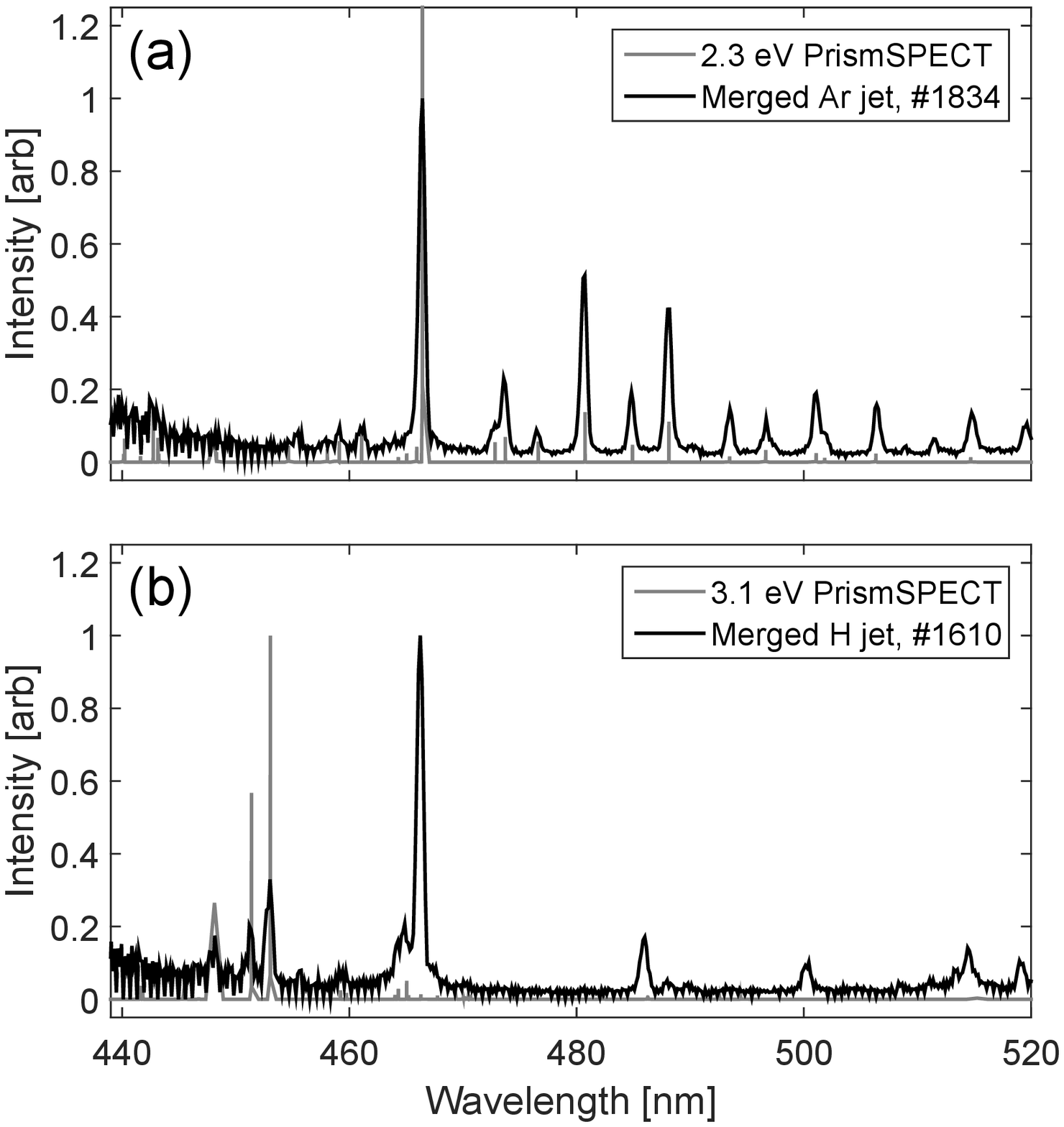}
	\caption{\label{fig:spec}Spectroscopic measurements for $t=35$~$\mu$s for (a) argon (shot \#1834) at $z=0$~cm [see Fig.~\ref{fig:time_series}(a) for spectrometer chord position] and (b) hydrogen (shot \#1610) at $z=-7.5$~cm [see Fig.~\ref{fig:time_series}(h) for spectrometer chord position].  Spectra calculated with PrismSPECT using the appropriate mixture, density, and temperature are also shown in each case.}
	\end{figure}

We calculate density using $\Delta\Phi=C_e (\bar{Z}-Err)\int n_{tot} d\ell$, where 
$C_e=\lambda e^2/4\pi\epsilon_{\text{0}}m_e c^2=1.58\times10^{-17}$~cm$^2$ is 
the phase sensitivity to electrons ($\lambda=561$~nm is the laser wavelength),
and $Err=0.08$ represents an upper bound on the phase sensitivity to ions.\cite{merritt14pop}  The $\bar{Z}$ is determined from spectrometer data and non-local-thermodynamic-equilibrium PrismSPECT\cite{macfarlane03, prism} calculations using the appropriate mixture (see Sec.~\ref{sec:impurities}).  
Lower bounds on peak $T_e$ and 
$\bar{Z}$ are inferred based on the appearance in PrismSPECT calculations of spectral lines seen in the data.
Upper bounds on peak $T_e$ and $\bar{Z}$ are inferred based on the appearance in the
calculations of spectral lines that are absent in the data. Examples of argon and hydrogen spectra plotted with PrismSPECT results are shown in Fig.~\ref{fig:spec}, and the details of spectral lines used to determine $T_e$ are given in Table~\ref{tab:spectral_data}. PrismSPECT calculations are density-dependent, and so the calculation of $n_{tot}$ and determination of $T_e$ and $\bar{Z}$ are iterated until self-consistent.\cite{merritt13prl,hsu12pop}
		
\begin{table}
\caption{\label{tab:spectral_data}Spectral lines used to determine $T_e$.}
\begin{ruledtabular}
	\begin{tabular}{l  c c c c }
    	Merging plasma & Ar & Ar  & H & H \\
    	Time ($\mu$s)& 35 & 40 & 20 & 35 \\
    	Density ($10^{14}$ cm$^{-3}$)& 1.5 & 2.5 & 0.81 & 2.1 \\   
    	Temperature (eV)& 2.3 & 2.8 & 1.4 & 3.1 \\ 	
    	Wavelength (nm)& 514.5 & 451.3 & 466.3 & 451.3  \\
    	Ion & Ar II & Al III & Al II & Al III  \\
    	Excitation potential\cite{NIST_ASD} (eV) & 19.5 & 20.6 & 13.3 & 20.6 \\
    	Transition probability\cite{NIST_ASD} ($10^{7}$ s$^{-1}$)& 1.06 & 20.9 & 5.81 & 20.9 \\
	\end{tabular}
\end{ruledtabular}
\end{table}	
		
For chord length, we use jet diameter divided by a factor of $\cos(30^{\circ})$ to account for the angle between the interferometer line of sight and the jet axis.  Jet diameter can be 
estimated in several ways: using a measured expansion rate---calculated from the difference in front and peak arrival 
time in interferometer data---together with the time-of-flight from nozzle to chamber center, using a theoretical expansion rate $v=2c_s/(\gamma-1)$, calculated with initial 
$c_s$ (where $c_s$ is the ion sound speed and $\gamma$ the ratio of specific heats) and
using an average mass based on mixture, together with the time-of-flight from nozzle to chamber center, or using full width at half maximum (FWHM) of visible emission in a CCD image lineout at the spectrometer 
chord position at $z=-7.5$~cm.  Table \ref{tab:diameter} presents diameters calculated by each method at $t=40$~$\mu$s for argon and $t=35$~$\mu$s for hydrogen; density calculated using the minimum and maximum diameter 
estimates differ by a factor of $\approx4$--$5$.
	
A direct measurement of electron density $n_e$
from Stark broadening in merging hydrogen 
jets at chamber center allows us to test the diameter estimates.  Using the measured 
$n_e=2\times10^{14}$ cm$^{-3}$ and $\Delta\Phi=7.3^{\circ}$ at $t=35$ $\mu$s, we infer a path 
length $\ell\approx40$~cm.  This implies a 35-cm jet diameter, consistent with the value from the fast-camera image lineout method.  Hence we use the fast-camera visible emission lineout FWHM to calculate $n_{tot}$.  The FWHM method also gives the smallest jet diameter estimates (Table \ref{tab:diameter}), giving a conservative upper limit on density and collisionality.
	
	\begin{table}
	\caption{Jet diameter (cm) using three estimation methods: using the theoretical expansion rate $2c_s/(\gamma-1)$ and time-of-flight, using expansion rate measured via interferometer and time-of-flight, and using the FWHM from a fast-camera image lineout.
	\label{tab:diameter}}
	\begin{ruledtabular}
	\begin{tabular}{l c c c}
    Species & Theory & Interferometer  & Image lineout \\ 
    \hline
    Argon & 130 & 29 & 24 \\  
    Hydrogen  & 93 & 100  & 34 \\  
	\end{tabular}
	\end{ruledtabular}
	\end{table}

	\subsection{Initial jet parameters}
\begin{figure}
	\includegraphics[width=\linewidth]{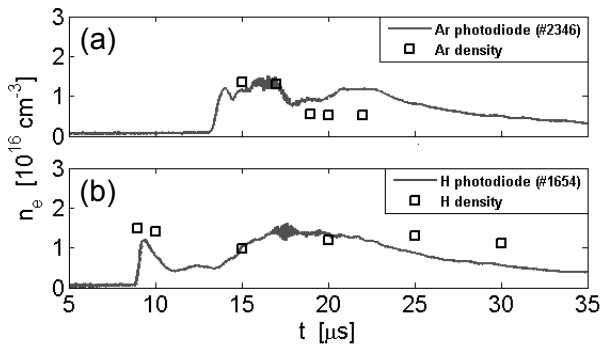}
	\caption{Measurements of electron density versus time at the gun nozzle show axial jet profiles for (a) argon and (b) hydrogen plasma jets.  Discrete density measurements from Stark broadening of H-$\beta$ (from multiple shots)
	overlay a continuous emission trace from the nozzle photodiode (scaled by the same factor in both cases).
	\label{fig:stark_trace}}
	\end{figure}

	Table \ref{tab:nozzle_param} gives plasma jet parameters measured at the gun nozzle ($z=111$ cm).
	Spectroscopy provides both $T_e$ and $n_{e}$ at the nozzle [Fig.~\ref{fig:time_series}(a)], $T_e$ by comparing with PrismSPECT calculations as described in Section \ref{sec:setup} and $n_e$ from Stark broadening of the 
	H-$\beta$ line.  Figure \ref{fig:stark_trace} shows electron density as a function of time for hydrogen and argon jets.  Peak
	$n_e=1.5\times10^{16}$ cm$^{-3}$ for hydrogen and peak $n_e=1.4\times10^{16}$ cm$^{-3}$ for argon.  
	Comparing hydrogen and argon photodiode data shows the difference 
	in the time of jet emergence from gun nozzle, with the more massive argon emerging at a later time.	
	
	\begin{table}
	\caption{Initial jet parameters at gun nozzle.\label{tab:nozzle_param}}
	\begin{ruledtabular}
	\begin{tabular}{l c c c c}
    Species & $n_{e}$ ($10^{16}$ cm$^{-3}$) & $T_{e}$ (eV) & $\bar{Z}$ & $v$ (km/s)\\ %
    \hline
        Argon  & 1.4 & 1.9 & 1.2 & 34\\   
        Hydrogen & 1.4 & 1.9 & 1.1 & 55\\   
	\end{tabular}
	\end{ruledtabular}
	\end{table}

\begin{figure}
	\includegraphics[width=\linewidth]{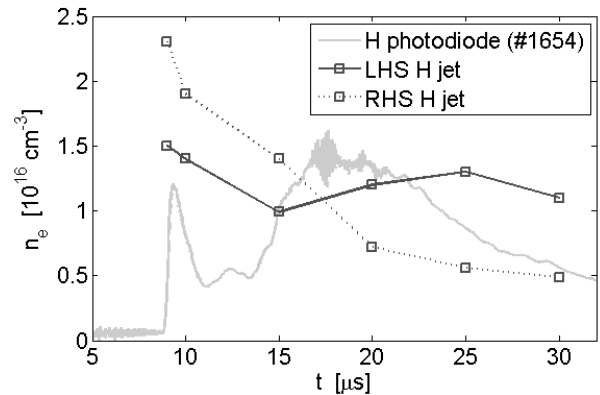}
	\caption{Stark-broadening measurements of $n_e$ at each gun nozzle show that although the instantaneous particle flow varies between the 
	two jets, the total particle input is roughly equal. \label{fig:jet_balance}}
	\end{figure}

	Stark-broadening measurements taken at the nozzle of each railgun for two sets of hydrogen experiments provide
	information on  the balance  between the two jets
	(Fig.~\ref{fig:jet_balance}).  The precise form of the density profile as a function of time differs between the 
	two, with density at any given time varying by a factor of 1.4--2.3 between the two jets, but the particle 
	input is comparable.  The total particle input, 
	as estimated using $v=34$ km/s from the right-hand-side jet photodiode traces for each jet and summing the total particle flux for 
	$t=9$--30~$\mu$s, is $N=8.6\times10^{17}$ for the right-hand-side jet and $N=7.3\times10^{17}$ for the
	left-hand-side jet.

	We expect $T_i\approx T_e$ before jet merging for both hydrogen and argon plasma jets, based on parameters measured at the gun nozzle (Table \ref{tab:nozzle_param}).  For argon, at $n_e=1.4\times10^{16}$~cm$^{-3}$, $T_{e}=1.9$ eV, $\bar{Z}=1.2$, using average mass $\bar{\mu}=28$ to account for mixture (described in Sec.~\ref{sec:impurities}) and 
	assuming $T_i\approx T_e$ (which provides the slowest equilibration), the 
	thermal-equilibration time is $\tau_{eq}=\bar{
	\mu} T_e^{3/2}/(3.2\times10^9 n_e Z^{2} \lambda_{ie})=0.2$ $\mu$s, where $\lambda_{ie}$ is the Coulomb logarithm.  For hydrogen, $n_e=1.4\times10^{16}$ 
	cm$^{-3}$, $T_{e}=1.9$ eV, $\bar{Z}=1.1$, and $\bar{\mu}=4.6$, which gives $\tau_{eq}=0.05$~$\mu$s.  The equilibration times are both shorter than the time for a jet to travel 1~cm.
	
The railgun imparts a magnetic field to the jet, measurable in the gun nozzle but which we expect to decay to negligible levels by the time of jet interaction.  Measurements of the resistively decaying field along the railgun nozzle give a decay time of 
$5.6$~$\mu$s.\cite{merritt14pop}  This predicts a magnetic field of $\sim1$~mT for argon and $\sim10$~mT for hydrogen by the time of jet-merging.  Estimated jet-diameter expansion by a factor of 4 for argon and 7 for hydrogen would drop the magnetic field strength to $\sim10^{-5}$~T in both cases.  A magnetic field probe near chamber center able to measure $\sim1$~mT fields measures no field in these jets; using an upper bound of 1~mT magnetic field strength, the ratio of jet kinetic energy density to magnetic energy density is $\sim 10^3$--$10^4$.  Hence, we treat the interactions as unmagnetized.      

\subsection{Impurities}\label{sec:impurities}
	We see indications of possibly high impurity levels in the plasma, based on
	the difference in pressure rise between neutral-gas-injection only and a 
	plasma shot in which neutral gas is injected and the railgun electrodes are discharged, as well as strong 
	impurity lines in spectroscopic measurements.  
	We estimate total impurity percentage using the difference in pressure rise; we take the ratio $\Delta P_{gas}/\Delta 
	P_{shot}$ to represent the overall percentage of working gas in the plasma experiment.
	Bright 
	aluminum and oxygen lines 
	suggest that impurities come primarily from the railgun insulator material, zirconia-toughened alumina 
	(0.85 Al$_2$O$_3$+0.15 ZrO$_2$).  We hence assume a relative percentage of aluminum and oxygen as plasma impurities based on their relative 
	percentage in the insulator, 61\% oxygen to 36\% aluminum. Impurity estimates do not account for the relative sensitivity of the pressure gauge to different 
	species, nor possible spatial or temporal 
	distribution of impurities within the plasma jet.
	
	Impurities are taken into account both in data analysis and in calculating collision scale lengths to determine plasma 
	collisionality.  To assess the sensitivity of the analysis to mixture, we performed analysis with bounding impurity percentages for argon, the species with the highest estimated percentage impurities.  We compared an upper limit of 60\% impurities, based on the chamber pressure difference, and a lower limit of 10\% 
impurities, chosen due to the appearance of impurity spectral lines at all times of interest, and we see that 
our collisionality-based physics conclusions are independent of the impurity 
percentage assumed within these bounds (see Table~\ref{tab:lengths}).  Considering argon mixtures with small amounts of carbon and hydrogen also gave no significant difference in results.  Inclusion of 1\% each carbon and hydrogen in both the 40\% and 90\% argon mixtures, and inclusion of 5\% each carbon and hydrogen in the 40\% argon 
	mixture, leaves $n$, $T_e$, and $\bar{Z}$ 
	unchanged.  The analysis assuming a 40\% argon, 60\% impurities mixture provides the most conservative collision lengths (i.e., shortest) and so will be used here.

Calculations of scale lengths for one-dimensional, two-fluid plasma shocks takes impurities into account via the use of the mixture-specific average ion mass, but 
any additional possible effects of impurities on shock dynamics are not addressed in this paper.

\subsection{Calculation of length scales}		
		We calculate inter-jet collision lengths for ion--ion, ion--electron, and 
electron--electron collisions, taking all ion species into account and calculating ion densities using each species percentage and ionization level.  Calculated lengths for the analysis in Sections \ref{sec:interpenetration} and \ref{sec:ionization} are given in Table \ref{tab:lengths} for argon and Table \ref{tab:H_lengths} for hydrogen. Calculations use relative velocity $v_{rel}$, determined by tracking the arrival time of the interferometer phase shift peak in each chord for single jet experiments.  

Electron--electron collision length is 
$\ell^{e-e}=v_{th,e}/\nu_e$, using thermal collision frequency $\nu_e$ because
the electron thermal velocity $v_{th,e}\gg v_{rel}$.  We calculate both slowing and perpendicular 
collision length scales for ion--ion and ion--electron interactions, using 
$\ell=v_{rel}/\nu$ for all cases---except ion--ion slowing length scale 
$\ell^{i-i'}_s=v_{rel}/4\nu_s$\cite{messer13pop}---where the slowing frequency 
$\nu_{s}$ and perpendicular collision frequency $\nu_{\perp}$ are calculated in 
the slow limit for ion--electron and the fast limit for ion--ion.\cite{remark} 
The total inter-jet ion--ion collision length for an ion species, taking 
interspecies collisions into account, is calculated by summing the collision 
frequencies for each collision type, e.g., for argon: 
$\ell_{s}^{\text{Ar}-i'}=
v_{rel}/\left[4 \,(\nu_{s}^{\text{Ar}-\text{Ar}}+
\nu_{s}^{\text{Ar}-\text{O}}+
\nu_{s}^{\text{Ar}-\text{Al}})\right]$.

\section{Interpenetration at initial jet-interaction}\label{sec:interpenetration}
\begin{figure}
		\includegraphics[width=\linewidth]{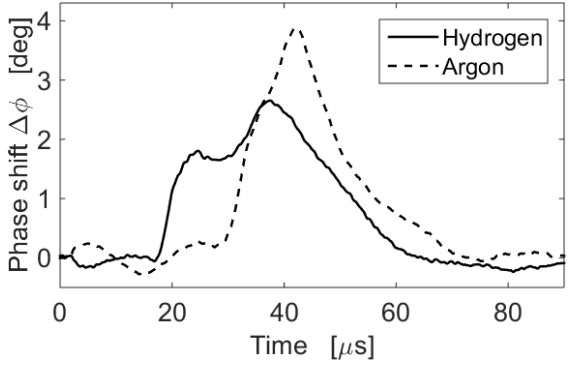}
		\caption{Average (over $\sim10$ shots) interferometer traces for RHS-only jet, measured at chamber center ($z=0$ cm).  Initial interaction time was determined by the arrival of the half-maximum: $t=35$~$\mu$s for argon and $t=20$~$\mu$s for hydrogen.  Peak arrival time is $t=40$~$\mu$s for argon and $t=35$~$\mu$s for hydrogen.  Late time was chosen to be a time at which the majority of the jet has arrived, but at which some plasma is still inflowing, $t=60$~$\mu$s for argon and $t=55$~$\mu$s for hydrogen.
		\label{fig:RHS_center}}
		\end{figure}

For both hydrogen and argon, three sets of interferometer measurements were taken: right-hand-side (RHS) jet only, left-hand-side (LHS) jet only, and two jets merging. 
Comparing $\Delta\Phi$ between
these three sets of experiments allows us to better understand the merging process and to quantify the jet interaction.  We estimate the time of intial interaction by the arrival of the single-jet $\Delta\Phi$ half-max at chamber center (Fig.~\ref{fig:RHS_center}).

\begin{figure}
		\includegraphics[width=\linewidth]{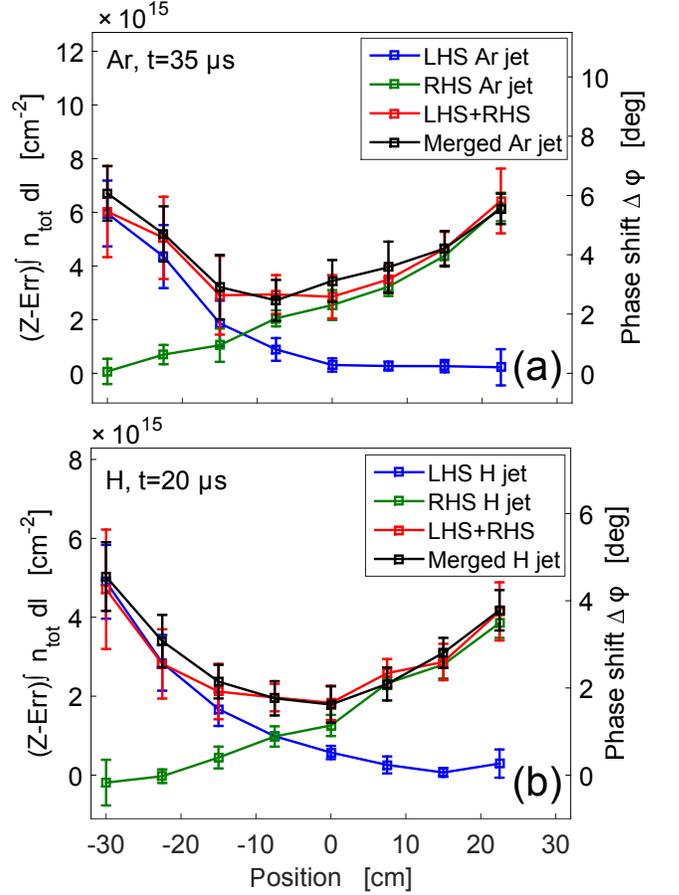}
		\caption{Interferometer traces indicate simple jet interpenetration at early time for both (a) argon and (b) hydrogen merging experiments.  Plots show spatial  profiles of $\Delta\Phi$ at $t=35$~$\mu$s for argon and $t=20$~$\mu$s for hydrogen for three sets of experiments: LHS jet only (blue trace), RHS jet only (green trace), and both jets merging (black trace).  The red trace is the sum of the LHS and RHS traces; the black trace overlays this at early times, consistent with simple interpenetration. Traces are averaged over $\sim10$ shots; error bars indicate standard deviation.   
		\label{fig:early_time_int}}
		\end{figure}

Figure \ref{fig:early_time_int} shows that the interaction at initial jet interaction is close to simple interpenetration for both argon [Fig.~\ref{fig:early_time_int}(a)] and hydrogen [Fig.~\ref{fig:early_time_int}(b)] merging experiments.  The plots show $\Delta\Phi$ at each chord location averaged over the $\sim 10$-shot data set for each of the LHS only, RHS only, and merging jet experiments.  The shape of the jet front is visible for the 
LHS only (blue trace) and RHS only (green trace) experiments; the sum of these two (red trace) represents the $\Delta\Phi$ expected in the case of simple jet interpenetration.  In both cases the merging-jet experiment values are very close to the sum of individual jets, indicating minimal interaction beyond simple interpenetration between the two jets.  

The observation of jet-interpenetration is consistent with calculated inter-jet collision lengths, which are much longer than the interaction region in the case of argon and larger than but of order the interaction length in hydrogen.  We note that these are conservative collision lengths:  single-jet interferometer traces 
indicate that both jets contribute to the total 
merged $\Delta\Phi$, making this an
upper bound on the density of 
each of the individual
interpenetrating plasma jets, and the leading edge interacting here is moving at a higher velocity (due to jet expansion)
than the jet bulk velocity, so the quoted 
velocities are a lower bound.  Density is overestimated and velocity underestimated, each of which decreases the calculated lengths, making the calculated lengths a lower bound.  Also, in the case of hydrogen, we can provide only an upper bound on $T_e$ and $\bar{Z}$; 
a lower $\bar{Z}$ would also increase calculated collision lengths.  

For argon experiments, the merged-jet $\Delta\Phi=3.1^{\circ}$ at $z=0$~cm, $t=35$~$\mu$s 
[Fig.~\ref{fig:early_time_int}(a)]. 
Using $\ell=20$~cm from the emission FWHM and $\bar{Z}=1.2$ from spectroscopic measurements and PrismSPECT calculations, 
this $\Delta\Phi$ corresponds to a density 
of $n_{tot}=1.5\times10^{14}$~cm$^{-3}$.  Using $v_{rel}=90$~km/s, $n_{tot}=1.5\times10^{14}$~cm$^{-3}$, $T_{e}=2.3$~eV, and $\bar{Z}=1.2$, we 
calculate inter-jet collision lengths as described above, presented in Table 
\ref{tab:lengths}.   Because the collision lengths are sensitive to $\bar{Z}$, we also determine the $\bar{Z}$ value corresponding to the upper and lower $\Delta\Phi$ values indicated by error bars in Fig.~\ref{fig:early_time_int}(a).  Both the upper and lower limits give the same value of $\bar{Z}=1.2$.  All ion collision lengths for $t=35$~$\mu$s are significantly longer 	
than the length scale of the experiment,
 consistent with the observation that the jets 
interpenetrate with minimal interaction.

In merging hydrogen, interferometer measurements give $\Delta\Phi=1.8^{\circ}$ at $z=-7.5$~cm, $t=20$~$\mu$s 
[Fig.~\ref{fig:early_time_int}(b)].  Spectroscopic measurements and PrismSPECT calculations indicate an electron temperature $T_e<1.4$~eV; we will use $T_e=1.3$~eV and the corresponding $\bar{Z}=0.71$ for length calculations.  As in the argon case, using the upper and lower limits of $\Delta\Phi$ leaves $\bar{Z}$ unchanged. Due to insufficient visible emission at $t=20$~$\mu$s, we use emission FWHM at $t=25$~$\mu$s to get $\ell=39$, and calculate $n_{tot}=8.1\times10^{13}$~cm$^{-3}$. We then use $v_{rel}=96$~km/s to calculate the inter-jet collision lengths presented in Table \ref{tab:H_lengths}.  Most ion collision lengths are significantly longer than the scale of the interaction, however the hydrogen stopping lengths are only slightly longer: 33--42~cm, while the jets have interpenetrated 15--30~cm.

\begin{table}
	\caption{\label{tab:lengths}Experimentally inferred plasma parameters and 
	calculated collision lengths for the 40\% argon mixture (90\% argon mixture 
	values, where different, are shown in parentheses).  Collisionality 
	estimates are qualitatively unchanged for the two mixture assumptions.}

	\begin{ruledtabular}
	\begin{tabular}{c c c c }
    	 & & t=35 $\mu$s & t=40 $\mu$s \\
    	\hline
    	$n_i$ ($10^{14}$~cm$^{-3}$) & & 1.5	(1.9) & 2.5 (2.4)		\\
    	$T_{e}$ (eV)			  & & 2.3 (1.7)	& 2.8	  \\
    	$\bar{Z}$				  & & 1.2 (1.0)	& 1.7 (1.8)	 \\
    	$Z_i$				& Ar	& 1.2	 (1.0) & 1.8 	\\
    						& O		& 1.0 		& 1.2	  \\
    						& Al	& 1.6	 (1.1) & 2.6  	\\					
    	\hline
		$\ell^{e-e}$ (cm) & 		& 0.054 (0.030)	& 0.035	(0.034)\\   	
 		\hline
     	$\ell_s^{i-e}$ (cm) &  Ar 	& 400 (360)	& 110		\\
     						&  O 	& 220 (140)	& 92 (90)	 	\\
     						&  Al 	& 160 (200)	& 37 (36)	 	\\	
     	$\ell_\perp^{i-e}$ (cm) &  Ar 	& $1.4~(1.8) \times10^{5}$  & 
			$3.2\times10^{4}$  \\
     							&  O 	& $3.3~(2.8)\times10^{4}$  & 
     		$1.1\times10^{4}$() 			  \\
     							&  Al 	& $3.9~(6.8) \times10^{4}$  & 
     		7500 (7400) 			  \\	
     	\hline
     	$\ell_s^{i-i'}$ (cm) &  Ar 	& 1000 (2000)	& 140 (170)		\\
     						 &  O 	& 390 (500)	& 81	(92)	\\
     						 &  Al 	& 350 (950)	& 41	(48)	\\	
     	$\ell_\perp^{i-i'}$ (cm) &  Ar 	& 5300 (8600)  & 710 (720)\\
     							 &  O 	& 1300 (1400)	& 260 	\\
     							 &  Al 	& 1400 (3300)	& 160 (170)   		\\	
		\hline
		Mach number & & 10 (12) & 8.4 (8.3)	\\
		\hline
		Interaction width (cm) & & 15--30 & 15--30	
	\end{tabular}
	\end{ruledtabular}
	\end{table}
	
	\begin{table}
	\caption{\label{tab:H_lengths}Experimentally inferred plasma parameters and 
	calculated collision lengths for the 80\% hydrogen mixture.}
	
		\begin{ruledtabular}
	\begin{tabular}{c c c c }
    	 & & t=20 $\mu$s & t=35 $\mu$s \\
    	\hline
    	$n_{tot}$ ($10^{14}$~cm$^{-3}$) & & 0.81	&  2.1		\\
    	$T_{e}$ (eV)			  & & 1.3	& 3.1  \\
    	$\bar{Z}$				  & & 0.71	& 1.2 \\
    	$Z_i$				& H		& 0.68	&  1.0	\\
    						& O		& 0.73 	& 1.3  \\
    						& Al	& 1.0	&  2.7	\\					
    	\hline
		$\ell^{e-e}$ (cm) & 		& 0.056	& 0.068\\   	
 		\hline
     	$\ell_s^{i-e}$ (cm) &  H 	& 42	&  	16	\\
     						&  O 	& 590	& 	 160	\\
     						&  Al 	& 550	& 	 68	\\	
		$\ell_\perp^{i-e}$ (cm) &  H 	& 780  & 120\\	
     							&  O 	& $1.7\times10^{5}$  & $1.9\times10^{4}$\\
     							&  Al 	& $2.8\times10^{5}$  & 	$1.4\times10^{4}$  \\	  	
     	\hline
     	$\ell_s^{i-i'}$ (cm) &  H 	& 33	& 1.8\\
     						 &  O 	& 990	&	40\\
     						 &  Al 	& 940	& 18\\	
     	$\ell_\perp^{i-i'}$ (cm) &  H 	& 110  & 5.5\\
     							 &  O 	& $2.3\times10^{4}$	& 750\\
     							 &  Al 	& $3.5\times10^{4}$	& 520\\	
		\hline
		Mach number & & 6.0 & 3.5\\
		\hline
		Interaction width (cm) & & 15--30 & 30--45	
	\end{tabular}
	\end{ruledtabular}
	\end{table}

\section{Increased mean ionization and decrease in inter-jet collision lengths}\label{sec:ionization}
Interferometer measurements taken at the time of arrival of the jet bulk show an increase in $\Delta\Phi$ over the sum of single-jet traces, indicating that the interaction can no longer be described as simple interpenetration.  In both argon and hydrogen experiments, the increase in $\Delta\Phi$ can be accounted for by an increase in
$\bar{Z}$ rather than an increase in density.

The $\Delta\Phi$ in argon jet-merging experiments at $z=0$~cm, $t=40$~$\mu$s 
is $\Delta\Phi=8.1^{\circ}$, for which our iterative process, using $\ell=22$~cm, gives 
$n_{tot}=2.5\times10^{14}$~cm$^{-3}$, $T_{e}=2.8$~eV, and $\bar{Z}=1.7$. 
Again, because the jets have interpenetrated, this 
bounds the density of the individual jets.  Here the upper-error-bar $\Delta\Phi$ value gives the same $\bar{Z}=1.7$, and the lower-error-bar $\Delta\Phi$ value gives $\bar{Z}=1.8$. Table 
\ref{tab:lengths} lists collision scale lengths calculated with $v_{rel}=90$~km/s and the $n_{tot}$, $T_{e}$, and $\bar{Z}$ values given above.  
The increase in $\Delta\Phi$ at $t=40$~$\mu$s in the merged-jet case over the simple-interpenetration case is consistent with a $\bar{Z}$ increase rather than a density increase.    
The inferred $\bar{Z}=1.7$--$1.8$ is a factor of $\approx 1.4$--$1.8$ increase from the $\bar{Z}=1.0$--$1.2$ expected for interpenetrating jets with no increased ionization (based on the $t=35$~$\mu$s measurement).  
The ratio of $\Delta\Phi$ in the merged-jet case to $\Delta\Phi$ for the sum of single jets at $z=0$~cm, $t=40$~$\mu$s is $8.1^{\circ}/5.9^{\circ}\approx1.4$; thus, increased ionization is sufficient to account for the increase in $\Delta\Phi$ between the two cases. 

In hydrogen experiments, the interferometer measures 
$\Delta\Phi=7.3^{\circ}$ at $z=-7.5$~cm, $t=35$~$\mu$s; using $\ell=35$~cm this gives 
$n_{tot}=2.1\times10^{14}$~cm$^{-3}$, $T_{e}=3.1$~eV, and $\bar{Z}=1.2$. The upper $\Delta\Phi$ value gives $\bar{Z}=1.1$, and the lower $\Delta\Phi$ value gives the same $\bar{Z}=1.2$. The ratio of the $\bar{Z}$ measured in the merged-jet case to that expected for interpenetration ($\bar{Z}=0.71$ measured at $t=20$~$\mu$s) is $1.2/0.71=1.7$.  The increase in merged-jet $\Delta\Phi$ at $t=35$~$\mu$s over the $\Delta\Phi$ expected in the case of interpenetration is $7.3/4.7=1.6$.  This indicates that ionization can account for the increase in $\Delta\Phi$ measured by the interferometer.  

The increase in $\bar{Z}$ measured in both argon and hydrogen experiments leads to a dramatic decrease in the calculated ion--ion collision lengths, which scale as $\bar{Z}^{-4}$.  The shortest inter-jet scale lengths in the argon mixture have now dropped to $\approx40$~cm (Table \ref{tab:lengths}), the scale of the jet interaction region.  The increase in $\bar{Z}$ is required to drop the collision lengths to the interaction scale length.  The density at $t=40$~$\mu$s with the lower $\bar{Z}$ value inferred at $t=35$~$\mu$s do not decrease the collision lengths to the interaction scale length.  In the hydrogen mixture, all inter-jet ion--ion collision lengths are now smaller than or on the scale of the interaction region (Table \ref{tab:lengths}).  In hydrogen experiments, the density at $t=35$~$\mu$s with the lower $\bar{Z}$ inferred at $t=20$~$\mu$s would be sufficient to drop the collision lengths to the interaction scale length.  However, we still observe that the increase in ionization precedes any increase in density.  Once the inter-jet collision lengths are small enough that each jet behaves as a barrier for the opposing jet, we expect plasma stagnation and a density increase.  

		\begin{figure}
		\includegraphics[width=\linewidth]{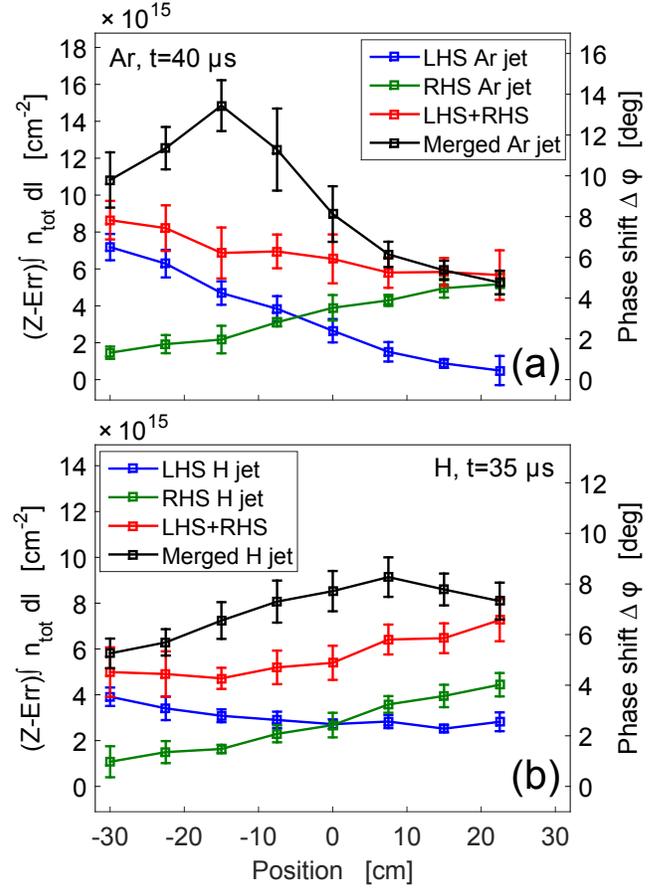}
		\caption{Interferometer traces show an increase in $\Delta\Phi$ attributable to increased ionization near the time of peak arrival for both (a) argon and (b) hydrogen merging experiments.  Plots show spatial profile of $\Delta\Phi$ at $t=40$~$\mu$s for argon and $t=35$~$\mu$s for hydrogen for three sets of experiments: LHS jet only (blue trace), RHS jet only (green trace), and both jets merging (black trace).  The red trace is the sum of the LHS and RHS traces; the increase in the black trace relative to the red trace is matched by the increase in measured $\bar{Z}$. Traces are averaged over 
		$\sim10$ shots; error bars indicate standard deviation.  \label{fig:mid_time_int}}
		\end{figure}

\section{Plasma stagnation and collisional shock analysis}
    
In both argon and hydrogen experiments, fast-camera images at late time (selected to be a time at which the majority of the jet has arrived, but at which there is still incoming plasma flow, see Fig.~\ref{fig:RHS_center}) show a well-defined emission structure (Fig.~\ref{fig:time_series}), and interferometer traces show a significant increase in $\Delta\Phi$ indicating density increase.  Interferometer measurements are consistent with plasma stagnation and with the formation of collisional shocks.

	\subsection{Argon shock analysis}
\begin{figure}
			\includegraphics[width=\linewidth]{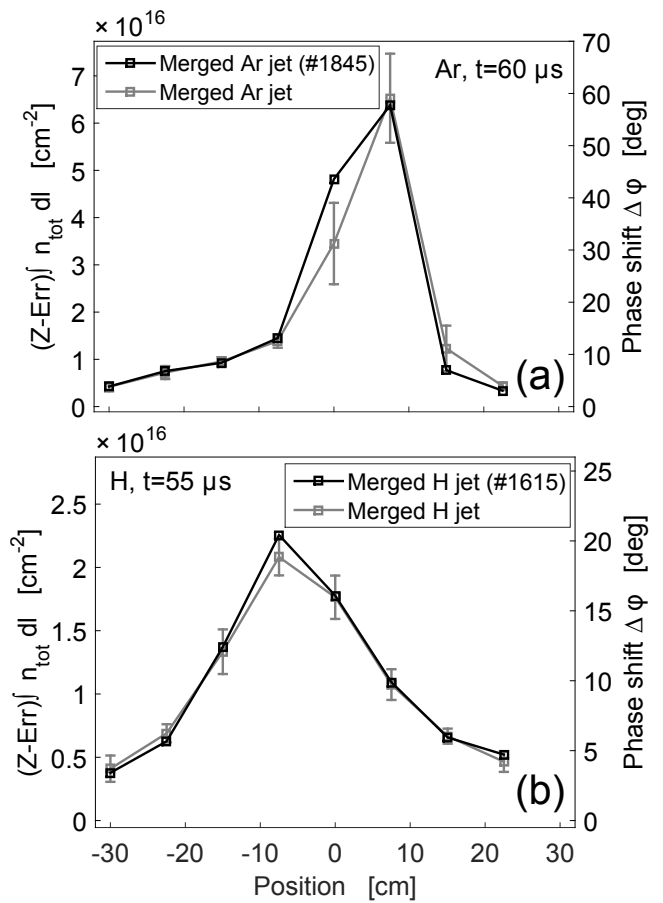}
		\caption{Interferometer traces show a pronounced peak, aligned with visible emission in fast camera images, due to plasma stagnation at late time for both (a) argon and (b) hydrogen merging experiments.  Plots show spatial profile of $\Delta\Phi$ at $t=60$~$\mu$s for argon and $t=50$~$\mu$s for hydrogen for the merging-jet data set average (gray trace) and for a single shot (black trace) corresponding to the fast camera image shown.  The scale-length of the drop in density is less than the inter-chord spacing, 7.5~cm.  Gray traces are averaged over 14 shots; error bars indicate standard deviation.
		\label{fig:late_time_int}}
		\end{figure}

Plasma stagnation leads to formation of a large, pronounced peak in the merged argon jet interferometer 
trace, with $\Delta\Phi_{peak}=59.1^{\circ}$, by $t=60$~$\mu$s.  Because shot-to-shot 
variation in $\Delta\Phi$ increases at later times, the interferometer data for 
the individual shot shown in the fast-camera image in 
Fig.~\ref{fig:time_series}(a) (\#1845) is plotted along with the 14-shot-averaged 
data in 
Fig.~\ref{fig:late_time_int}(a).  For the individual trace, a pronounced peak 
spanning two interferometer chords drops to $1/8$--$1/3$ the peak value on 
either side ($\Delta\Phi=43.6^{\circ}$ and $57.7^{\circ}$ to 
$\Delta\Phi=13.1^{\circ}$ and $7.0^{\circ}$).  This peak aligns with the 
region of increased emission in Fig.~\ref{fig:time_series}(a).  The chord corresponding to the spectrometer view is near the 
peak edge and measures $\Delta\Phi=43.6^{\circ}$; with $\ell=29$~cm this corresponds 
to $n_{tot}=1.3\times10^{15}$~cm$^{-3}$, $T_{e}=2.2$~eV, and $\bar{Z}=1.4$.  This may underestimate the actual diameter of the plasma, and so we also perform the analysis using $\ell=44$~cm, determined from the full-width-10\%-maximum from the camera lineout.  This gives $n_{tot}=6.6\times10^{14}$~cm$^{-3}$, $T_{e}=2.4$~eV, and $\bar{Z}=1.5$.
	
	The observed density transition scale is $\leq 7.5$~cm [see Fig.~\ref{fig:late_time_int}(a)], the 
	inter-chord spacing.  We can compare this to the predicted shock thickness, which is of order the 
	post-shock ion mean-free-path $\lambda_{mfp,i}=v_{th,i}/\nu_i$, \cite{jaffrin64pof} where 
	$v_{th,i}$ and $\nu_i$ are the shocked-ion thermal velocity and collision rate, respectively.

	A one-dimensional, two-fluid plasma shock model,\cite{jaffrin64pof} generalized to non-hydrogenic 
	species, allows prediction of post-shock $T_i$ from the plasma parameters measured at 
	$z=0$~cm.  An ideal monatomic gas has adiabatic index $\gamma=5/3$; however, we expect a 
	reduced $\gamma$ due to ionization and excitation. \cite{zeldovich}  
	We use PROPACEOS\cite{prism} to perform equation-of-state calculations in the pre-shock plasma 
	parameter range using values measured at $t=40$~$\mu$s, and then determine $\gamma$ using $\gamma = P/\rho\mathcal{E}+1$, where $
	\mathcal{E}$ is internal energy, $P=n_e k T_e+n_i k T_i$ is pressure, and $\rho$ is mass 
	density.\cite{zeldovich} For the 40\% argon mixture we use $n_i=2.5\times10^{14}$~cm$^{-3}$, 
	$n_e=\bar{Z} n_i$, 
	and $T_i=T_e=2.8$~eV, and we calculate $\gamma=1.2$.  Solving jump conditions for these pre-shock 
	values, assuming that $T_e$ does not change across the shock,\cite{jaffrin64pof} and using $\bar{\mu}=27$ and
	$v_{rel}=45$~km/s between the plasma jet and the stagnated plasma predicts post-shock $T_i=61$~eV for both sets of values calculated above.  The corresponding shock width is of order 
	$\lambda_{mfp,i}\approx2.4$--$3.4$~cm.

\subsection{Hydrogen shock analysis}
	Hydrogen interferometer traces [Fig.~\ref{fig:late_time_int}(b)] also show a peak due to plasma stagnation at late time ($t=55$~$\mu$s), though it is less pronounced than the argon case ($\Delta\phi=20.4^{\circ}$ compared to $\Delta\phi=59.1^{\circ}$).  In Fig.~\ref{fig:late_time_int}(b) we again we show both the data for 
the individual shot shown in the fast-camera image in 
Fig.~\ref{fig:time_series}(l) (\#1615) and the 14-shot-averaged 
data. The density drop from peak to outside the peak can be bounded only by twice the interfermeter chord spacing, from the peak at $z=-7.5$~cm to $z=-22.5$~cm [Fig.~\ref{fig:late_time_int}(b)], though we note that even a very short density transition scale would produce this measurement if centered on the chord at $z=-15$~cm.   Fast-camera images show a similarly sharp structure [Fig.~\ref{fig:time_series}(l)], though the bounds of the visible emission do not correspond as well to the interferometer peak as in the argon case.

The single-shot merged-jet phase 
	shift at $z=-7.5$~cm is $\Delta\Phi=20.4^{\circ}$, which gives 
	$n_{tot}=8.2\times10^{14}$~cm$^{-3}$, $T_e=2.5$, and $\bar{Z}=1.1$ for $\ell=27$~cm.  Instead using full-width-10\%-maximum to determine $\ell=51$~cm gives $n_{tot}=4.3\times10^{14}$~cm$^{-3}$, $T_e=2.7$, and $\bar{Z}=1.1$. We again use measured values and the one-dimensional, two-fluid jump conditions to determine post-shock $T_i$.  For the measured post-shock values above, assuming $T_e$ is constant across the shock and using $\gamma=5/3$, $\bar{\mu}=4.6$, unshocked $\bar{Z}=1.1$ and $v_{rel}=48$~km/s gives post-shock $T_i=42$~eV for both cases.  This corresponds to a shock thickness of $\lambda_{mfp,i}\approx4.2$--$7.6$~cm, less than or of order the inter-chord spacing, indicating that the bounded density gradient scale length is consistent with a collisional shock.

\section{Summary and discussion}
Space- and time-resolved measurements of the head-on merging of two supersonic
plasma jets show a progression from  
interpenetration to increased ionization and finally to stagnation and density amplification consistent with collisional shock formation. Inter-jet collision 
lengths are greater than the interaction scale 
in the interpenetration stage, consistent with simple interpenetration seen in interferometer measurements. 
In the ionization phase, we measure increased $T_e$ and $\bar{Z}$.  
Calculated ion--electron slowing lengths, the shortest of which are of the order of the interaction region length for both argon and hydrogen, suggest that frictional heating of electrons by ions of the opposing jet accounts for the increase in $T_e$. 
Based on available ionization rate coefficient data (e.g., \textcite{chung05}), it is unlikely that
electron-ion impact ionization can account for the rise in $\bar{Z}$ during the ionization phase (over $\sim 5$~$\mu$s), nor is ion-impact ionization between the counter-streaming ions likely.\cite{hasted72}
Further work is needed to identify the exact mechanism of the
$\bar{Z}$ increase on the observed time scale.  Based on theoretical calculations for a hydrogen plasma\cite{stringer64} and qualitative extrapolations to
non-hydrogenic ions (detailed calculations for the latter should be done in future work), we note that two-stream instabilities are unlikely to play a role in our experiments because our counter-streaming velocities are outside the range needed for electron--electron, electron--ion, as well as ion--ion instabilities.
The increased $\bar{Z}$ affects inter-jet ion slowing lengths, which decrease to the interaction region width, and the merging jets begin to stagnate.  
This increase in $\bar{Z}$ mediates the transition from collisionless interpenetration to collisional stagnation in both argon and hydrogen experiments, and is required for the transition in the argon case.
Once the jets stagnate, a 
bright structure visible in fast-camera images appears, corresponding to a region of 
increased density measured by the interferometer.  The transition between high 
and low density occurs on a length scale less than the interferometer inter-chord spacing of 
$7.5$~cm in argon and less than twice this spacing in hydrogen; the approximate shock thickness expected from one-dimensional, two-fluid estimates are a few centimeters, consistent with these observed transition lengths.  This is a concrete example of colliding supersonic plasmas transitioning from a collisionless to collisional interaction owing to the effects of increased $\bar{Z}$.  The data presented here
can aid in the validation of plasma collisionality and ionization models in the presence
of complex equation of state\@.

\acknowledgements{We thank John Dunn, Colin Adams, and Elizabeth Merritt for technical assistance, and
Colin Adams and Igor Golovkin for useful discussions.  This work was sponsored by the Laboratory Directed
Research and Development (LDRD) Program of LANL under DOE contract no.\ DE-AC52-06NA25396.}

%

\end{document}